# The Development of Microfluidic Systems within the Harrison Research Team

*D. Jed Harrison - University of Alberta / National Institute for Nanotechnology*

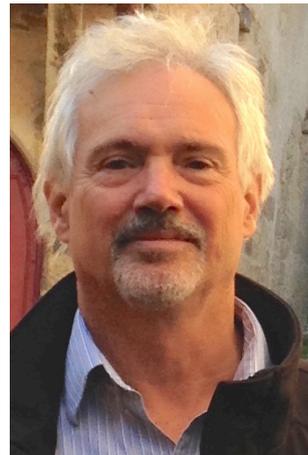

## Biography

D. Jed Harrison is a Full Professor of the Department of Chemistry at the University of Alberta, Canada. He began as an Assistant Professor in 1984, was promoted to Full Professor in 1994, and served as Chairman of the Department of Chemistry from 2006-16. He is a member of the Canadian National Research Council's National Institute for Nanotechnology. He received his B.Sc. in Chemical Physics from Simon Fraser University and earned a PhD in Chemistry at the Massachusetts Institute of Technology. His research interests are generally in the application of microfabrication methods and micromachining technologies to analytical systems and chemical sensors. Harrison is the co-author of over 225 scientific publications, and works with a research group of undergraduate, graduate and Post-Doctoral students that numbers about 10. He was the founding President of the Chemical and Biological Microsystems Society, and a key organizer of the Micro Total Analysis Conference in MicroFluidics from 1996 to 2009. He is the recipient of a number of awards, including the Canadian Society for Chemistry's McBryde Medal and Maxxam Award, the NSERC Steacie Memorial Fellowship, the Heinrich Emanuel Merck Prize, the Golay Award and the American Chemical Society's Instrumentation Award. He is an elected Fellow of the Royal Society of Canada.

## Abstract

My research team's initial entry into microfluidics involved the first demonstration of an integrated separation system for samples in liquids, based upon electrokinetic flow and electrophoretic separation of ions. Our first work in this area, done jointly with Andreas Manz and his team [1-5], developed methods to fabricate devices in glass, along the way solving the problem of bonding glass devices [2-4]. The bonding methods were key elements of later technology transfer to assist Caliper and Micralyne in their commercialization efforts. We demonstrated that high quality electrophoretic separations could be achieved on chip [2-5]. The application of very high electric fields and very rapid separation times was readily achieved as a result of design properties that could be readily fabricated in a planar format, Fig 1. This was, for example, noted in a 1992 presentation, "*Manipulation of the channel geometry to control where the bulk of an applied potential drops was shown to be easily accomplished. High electric fields of up to 2500 V/cm were sustained within the channels, exceeding the value of 300 V/cm typically used in conventional capillaries*" [5].

We also demonstrated that flow at channel intersections and junctions could be controlled by voltages. Voltage driven flow allowed mixing, proportioning of flow, reversing flow at a junction by reversing fields, and preventing leakage from side channels [6-8]. Controlled flow allowed the integration of injector designs, and of mixing of reagent streams without the need for integrated moving parts, and was a key step in the early development of microfluidics, when satisfactory valving technologies were not available. As stated at Transducers 93 [8], "*This result also clearly shows that control of the potential of all of the channels could be used to control the leakage phenomenon we have observed previously. More importantly, these results show that a common sample pre-treatment step, dilution, can be effected within the capillary channel manifold.*"

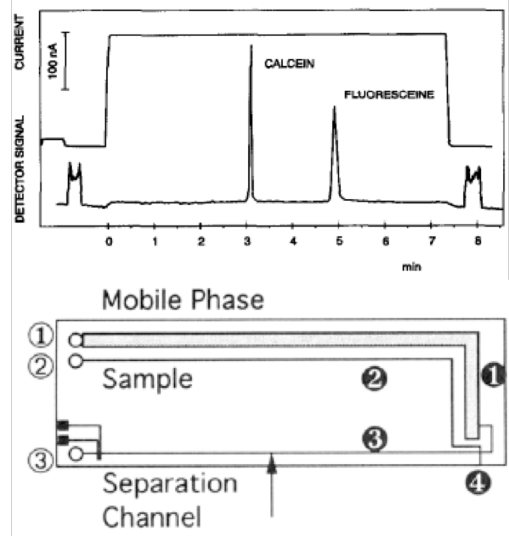

Fig. 1. First capillary electrophoresis separation on-chip, ref 2-5.

These ideas of valveless control of flow, reagent mixing, and rapid separation were then brought together and clearly demonstrated as integrated analytical tools [6].

Our team then realized the concepts of mixing reagents, reacting them, and separating the products for chemical analyses through a series of studies on immunoassays, DNA hybridization and linear, isothermal DNA amplification techniques [9-17]. The use of both post-column and pre-column reactions with electrically metered mixing and electrophoretic separations was demonstrated [9-11]. This work illustrated that the phrase we first coined, "a laboratory on a chip" [5], was a realistic goal.

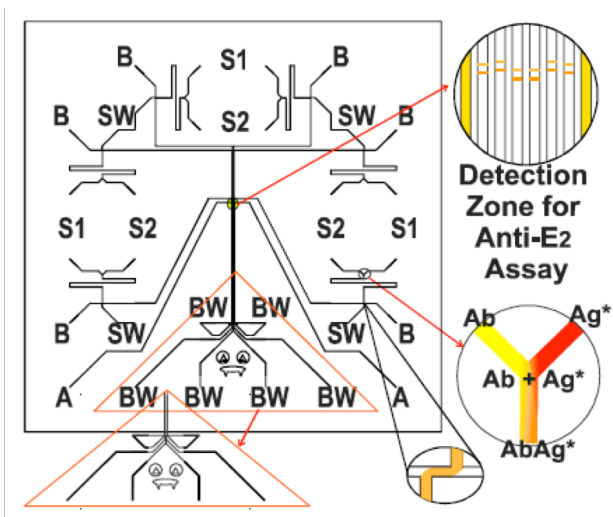

Fig. 2. 6-channel mix, react, electrophoesis separation device for multiplexed immunoassays, ref 16.

My team paid particular attention to applications in clinical diagnostics using immunoassays on chip, in which sample was metered, injected, mixed with reagents, separated and analyzed in complex integrated and multiplexed devices [10, 12-17]. We demonstrated that quantitative assays involving each of these integrated steps could be readily performed. Work on assembling a complete portable system capable of performing immunoassays [15] led to the commercialization of the Micralyne MicroFluidics Toolkit. These efforts culminated in the design and demonstration of a multiplexed device capable of simultaneous multiple independent immunoassays, for simultaneous analysis of multiple samples, Fig 2, or simultaneous self calibration and sample analysis [16].

Integrating the development of a series of standard dilutions to allow internal, multi-point calibration was also developed [17].

A linear, isothermal DNA amplification technology was selected for demonstration of on-chip mix, react and separate strategies for diagnostic molecular biology analysis. Sample, DNA templates and enzymes were metered and mixed on-chip, delivered to a heated zone, and then an electrophoretic DNA separation was run to analyze for product in a quantitative assay [18-19].

Detection in a microchip environment was regarded as a challenge early on, due to the small volumes and path lengths involved. We initiated a series of developments in optical detection methods. We were the first to show a number of techniques on chip such as fluorescence, absorbance and chemiluminescence could give the same concentration detection limits as in conventional formats, laying to rest this consistent early criticism [20-23].

Each of the above studies was focused on demonstrating the utility of the mix, react and separate strategy within an integrated device environment, and to illustrate that integration of multiple laboratory functions was feasible and beneficial.

Microfluidic devices offer unique opportunities for the study and analysis of cells, in particular studies of single cells. The volume of the channels and area surrounding a cell are well matched to the internal volume of a cell, so that dilution of internal cell components upon their release is minimized, and the released components are readily accessed within the channels for analysis. We entered this field with a demonstration that electrokinetic forces can be used to control the flow of cells, and manipulate the flow of individual cells at intersections, without wholesale cell damage [24, 25]. This result was followed by analyses of single cell cell contents, released by chemical lysis [26, 27], illustrating the small volume environment could be used effectively for analysis of specific enzymes within a cell. We also showed that mixing within these small volumes was efficient, allowing evaluation of the kinetics of the response of single cells to treatment with drugs and other antagonist or protagonist agents [28].

Processing of larger numbers of cells on chip is also of clear significance, and our research demonstrated the ability to electroporate and transfect significant quantities of cells on chip, emulating the performance of conventional systems [29]. The quantities generated were sufficient for conventional agar plate, anti-bacterial incubation techniques, unlike other micro-devices that demonstrated electroporation of only a small number of cells at a time. We also were the first to illustrate that cell rolling studies could be implemented on chip to evaluate the impact of various drugs on cell adhesion, greatly reducing the demand for the often expensive reagents required in such studies [30].

Mass spectrometry (MS) provides a very powerful analytical tool, and the internal volume requirements of MS are very well matched to the volume within a microfluidic system. We developed the first ultra-low dead volume electrospray interface between a separation chip and the MS [31, 32]. Combining the mix, react, separate device designs described above with a low dead volume interface allowed for shortened and very effective analysis of proteins from cell lysates, including the challenging membrane proteins, for applications in Proteomics. Packed bed reactors were added to these systems to create rapid tryptic digestion of proteins for protein identification and analysis of post-translational modifications [33, 34]. This work was followed by extensive efforts to develop multiplexed analysis systems within electrokinetic systems for separation, sample fractionation, collection and enrichment, with detection by MS [35-38]. In addition to the Proteomics applications, the work illustrated the complexity and sequencing of process and flow schemes that can be achieved, based upon the initial developments of electrokinetic systems in planar microfluidic systems. Our research team was the first to demonstrate that bead beds could be effectively incorporated into microchips, Fig 3 [39-41] for analytical purposes. This approach has real advantages compared to in-situ fabrication of separation or reactions beds, since commercially available, reproducible beads can be utilized to obtain high reproducibility. Bead beds were developed for reaction beds, solid phase extraction (SPE), ion exchange and reverse phase electrokinetic chromatography (CEC), and this technology has been used extensively since that time in a large range of applications [33, 34, 39-41]. More recently we developed methods to fabricate crystalline packed beds, with very low defect density for DNA and protein separation [42-44]. This is a very powerful technology that has allowed 50-100 nm plate heights to be achieved for separation of SDS denatured proteins with sub-kilo-Dalton mass resolution [45].

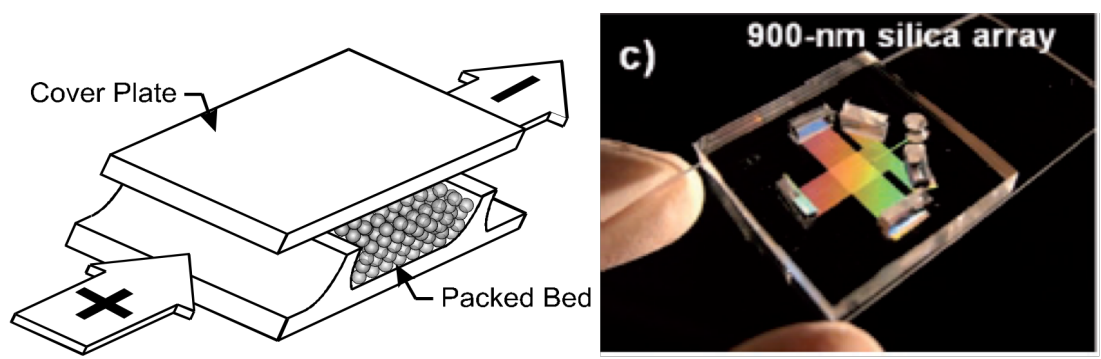

*Fig. 3 left, Schematic of packed bed design for SPE, CEC, reaction beds; right, photo of crystalline particle bed for DNA separation, ref 39-44.*

In retrospect, it long ago became clear that the primary reason that electrokinetic pumping technology defined the early versions of microfluidics, and took off so substantially, was because it did not require integrated pumps and valves. The early pumps and valves in non-compliant silicon materials simply did not work well enough to be the foundation of an industrial technology. The fabrication of interconnecting flow channels in the valveless system afforded by electrokinetics is simple, allowing sophisticated integration of analytical processes, kick starting lab on a chip concepts into real devices. These systems continue to be explored today, and are available commercially.

## Acknowledgements

Harrison and his team are grateful to many funding sources over the period of this research effort, including Canadian agencies NSERC, NRC, NINT, Genome Canada, CFI, AHFMR, AIHS, NanoBridge, the University of Alberta, the USA's DARPA, and companies including Ciba-Geigy, MDS-Sciex, ABI-Perkin Elmer, Shimadzu, Affymax and Caliper. Harrison gratefully thanks all of his team members and colleagues over the years, those who are specifically cited below and those who are not.